\documentclass[12pt,preprint]{aastex}

\shorttitle{Orion A and the ONC}
\shortauthors{Hartmann and Burkert}

\begin{document}

\title{On the Structure of the Orion A Cloud and the Formation of the Orion Nebula Cluster}

\author{Lee Hartmann\altaffilmark{1} and Andreas Burkert\altaffilmark{2}}

\altaffiltext{1}{Dept. of Astronomy, University of Michigan, 500 Church
Street, Ann Arbor, MI 48109}
\altaffiltext{2}{University Observatory Munich, Scheinerstrasse 1, D-81679 Munich, 
Germany}

\email{lhartm@umich.edu, burkert@usm.uni-muenchen.de}

\newcommand\msun{\rm M_{\odot}}
\newcommand\lsun{\rm L_{\odot}}
\newcommand\msunyr{\rm M_{\odot}\,yr^{-1}}
\newcommand\be{\begin{equation}}
\newcommand\en{\end{equation}}
\newcommand\cm{\rm cm}
\newcommand\kms{\rm{\, km \, s^{-1}}}
\newcommand\K{\rm K}
\newcommand\etal{{\rm et al}.\ }
\newcommand\sd{\partial}

\begin{abstract}
We suggest that the Orion A cloud is gravitationally collapsing on
large scales, and is producing the Orion Nebula Cluster due to 
the focusing effects of gravity acting within a finite cloud geometry.  
In support of this suggestion,
we show how an elliptical rotating sheet of gas with a modest density 
gradient along the major axis can collapse to produce a structure 
qualitatively resembling Orion A, with a fan-shaped structure at
one end, ridges or filaments along the fan, and a narrow curved filament
at the other end reminiscent of the famous integral-shaped filament.  
The model produces a local concentration of mass within the narrow filament 
which in principle could form a dense cluster of stars
like that of the Orion Nebula.  We suggest that global
gravitational contraction might be a more common feature of molecular
clouds than previously recognized, and that the formation of 
star clusters is a dynamic process resulting from the focusing effects
of gravity acting upon the geometry of finite clouds.
\end{abstract}


\section{Introduction}

The dynamical state of molecular clouds clearly plays an important role in 
star formation.  Older concepts of long-lived clouds, maintained
in rough equilibrium by balancing turbulence and magnetic forces against
gravity, are being challenged by theoretical results suggesting rapid dissipation
of magnetohydrodynamic turbulence which could lead to collapse 
(e.g. Mac Low \etal 1998; Stone, Ostriker, \& Gammie 1998; Ostriker, Gammie, \& Stone 1999) 
as well as observational constraints from
stellar populations which indicate short cloud lifetimes (Ballesteros-Paredes,
Hartmann, \& Vazquez-Semadeni 1999; Elmegreen 2000;
Hartmann, Ballesteros-Paredes, \& Bergin 2001; see Ballesteros-Paredes \etal 2006
and Burkert 2006 for a recent reviews).  In addition, we recently pointed out 
the difficulties in setting up a velocity field which would globally stabilize a 
finite cloud containing many Jeans masses against its self-gravity
(Burkert \& Hartmann 2004; BH04); our results suggested 
that molecular clouds of appreciable mass are likely to 
be collapsing, at least in some regions.  These developments suggest 
that star-forming molecular clouds are relatively transient, rapidly evolving objects.

The dynamics of molecular clouds are also likely to be highly relevant
to the formation of clusters, in which most newly-formed stars reside
(e.g., Lada \& Lada 2003). 
A number of dynamical simulations of star formation have been carried out
(e.g., Klessen \& Burkert 2000, 2001; Bate, Bonnell, \& Bromm 2002, 2003),
with some semi-analytic investigations (e.g., Krumholz, McKee, \& Klein 2005),
but the processes responsible for assembling the protocluster gas have
been given much less attention; for example, in the numerical work of 
Bonnell, Bate, \& Vine (2003), the protocluster aggregation of gas 
is simply assumed as an initial condition.  In BH04 we suggested that 
the assemblage of protocluster gas might be due to the focusing effects of
gravity in finite, collapsing clouds.

The Orion A molecular cloud is a favorable site for studying the relationship
between cloud dynamics and cluster formation, as it contains
the Orion Nebula Cluster (ONC), the closest, relatively populous 
($\gtrsim 2000$ members) young cluster containing relatively massive
(O, early B) stars.  In BH04 we presented 
an extremely simplified model of the gravitational collapse of an elliptical sheet
which exhibited features in rough qualitative agreement with the structure
of the Orion A cloud, forming a mass concentration near one end of the cloud which
might represent the accumulation of gas for a large cluster like the ONC.
In this paper we present a modified model of sheet collapse which exhibits
a remarkable similarity to the structural features of the Orion A cloud. 
This picture makes predictions for the expected proper motions of the stellar
members and suggests new perspectives both on the nature of "turbulence" in the region 
and on the conditions under which star clusters form.

\section{Orion A and the ONC}

A huge literature has been devoted to Orion A and the ONC (see 
Genzel \& Stutzki 1989, O'Dell 2001, and references therein).
Here we review the large scale morphology and kinematics of the dense
gas in the cloud and the spatial relationship to the ONC.

\begin{figure}
\plotone{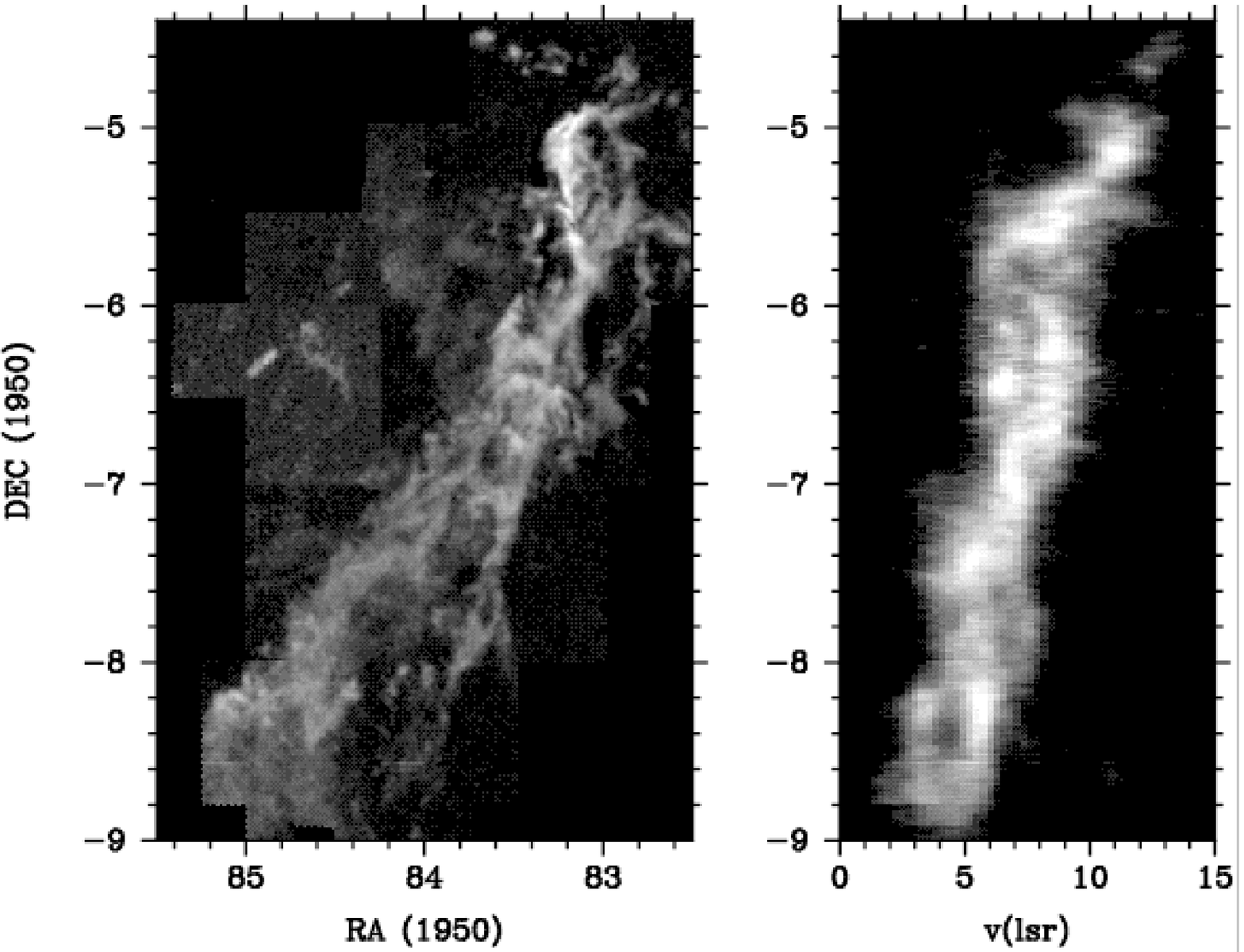}
\caption{Large-scale $^{13}$CO distribution in the Orion A cloud, 
taken from Bally \etal (1987).  The left panel shows the total
brightness distribution; the overall ``V''-shaped nature of the cloud is
apparent, with the famous ``integral-shaped'' filament apparent at
the northern end of the cloud.  The right panel shows the velocity 
distribution of the $^{13}$CO emission, summed over right ascension, as 
a function of declination (see text)
}
\label{fig:w13big}
\end{figure}

Figure \ref{fig:w13big} shows the large-scale $^{13}$CO brightness map of the Orion
A cloud from Bally \etal (1987).  The left panel shows the total surface brightness
map, with the well-known narrowing of the cloud proceeding from south to north
in roughly a ``V'', culminating in the famous integral-shaped filament at the northern
end.  The ONC resides near the middle of the integral-shaped
filament (see below). 

The right panel of Figure \ref{fig:w13big} shows the radial velocity 
distribution of the $^{13}$CO emission
summed over right ascension as a function of declination.  Here we call attention
to features already discussed by Kutner \etal (1977),
Bally \etal (1987), Maddalena \etal (1986), and Heyer \etal (1992) but which deserve particular
emphasis.  First, there is the well-known overall radial velocity gradient from south to
north, with a velocity difference of $\sim 8 \kms$ from one end to the 
other (see also Wilson \etal 2005).  
If we take the total mass of the Orion A cloud to be
$1.1 \times 10^5 \msun$ from Wilson \etal (2005), and a half-cloud length $R$ of 2.5 degrees
$\sim 16$~pc at a distance of 470 pc, then a characteristic virial velocity is
$(GM/R)^{1/2} = 4.9 \kms$.  Thus, the observed $\sim \pm 4 \kms$ systematic
velocity gradient along the cloud is dynamically significant (Kutner \etal 1977; 
Bally \etal 1987). 

The second general feature worth noting in the right panel of
Figure \ref{fig:w13big} is that the velocity distribution
shows many coherent structures (e.g., Bally \etal 1987).  Significant lumps of gas with
distinct kinematics can be seen all along the cloud, with velocity differences
of roughly $2-3 \kms$.  Heyer \etal (1992) found that some of these structures
appear to be bubbles in $^{12}$CO, and suggested that they might be wind-driven.
There is a particularly large velocity systematic gradient at the 
northern end of the cloud running from a velocity of $11 \kms$ to about $7-8 \kms$ 
proceeding southward, corresponding to the integral-shaped filament (see also Wilson \etal 2005). 
There is also evidence for an especially large
local velocity dispersion at a declination of $\sim -5.4^{\circ}$ (1950), which
is near the center of the ONC.

To illustrate the spatial distribution of the young stars in and near the ONC 
we use infrared-detected stars selected from the systematic survey of Carpenter,
Hillenbrand, \& Skrutskie 
(2001) from deep scans using the 2MASS telescope.  Although it is impossible to
exclude non-members completely from this data, it is useful for our limited purposes
because it is not optically-biased (although extinction can still have some effect)
and has relatively uniform sensitivity over a wide area.  
To limit the contamination by non-members, we used the pre-main sequence tracks
in the $K$ vs. $H-K$ diagram as developed by Hillenbrand \& Carpenter (2000)
for a 1 Myr-old population at 480 pc.  Following this prescription,
we included all stars with $K_s \leq 10.5$, plus
all stars with $10.5 < K_s \leq 12.5$ and $H-K_s \geq 0.4 + 0.1(K_s -13.0)$.
These selection criteria probably do not remove all non-members but yield a relatively
unbiased, large-scale view of the stellar population.

\begin{figure}
\epsscale{1.0}
\plotone{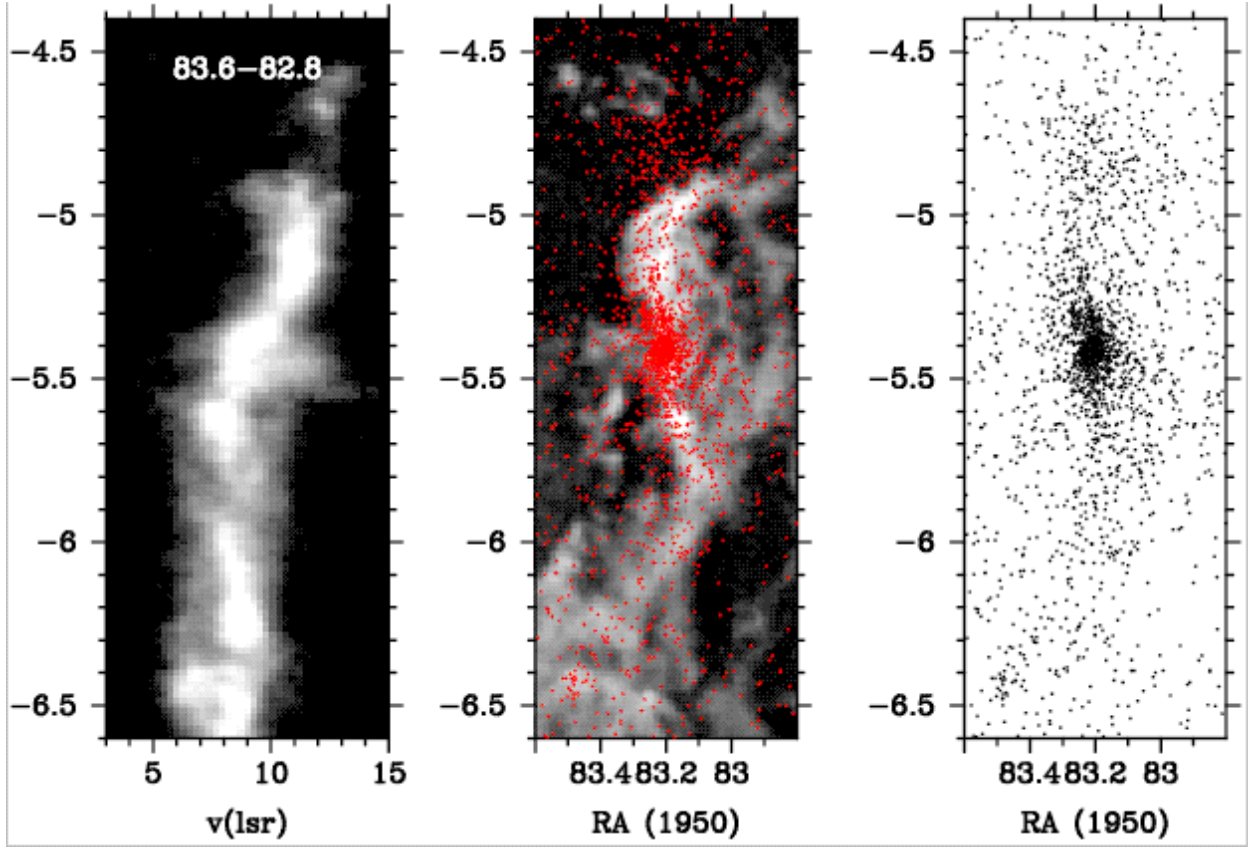}
\caption{Spatial distribution of stars and gas in Orion A, compared with local kinematics.
Left panel: $^{13}$CO velocity-declination plot between
RA $= 83.6$ and 82.8, taken from the data set of Bally \etal (1987).
Right panel: distribution of near-infrared selected stars from Carpenter \etal (2001)
(see text).  Middle panel: superposition of stars onto the total $^{13}$CO emission.
.}
\label{fig:w13bigstars}
\end{figure}

In Figure \ref{fig:w13bigstars} we show a closer view of the ONC region.
The right panel shows the spatial distribution of the infrared sources as selected
above; this is very similar to that found by Carpenter \etal (2001) for infrared
variables  (likely pre-main sequence members), though this selection resulted in 
fewer stars given photometric uncertainties and small amplitudes of variability in
the near-infrared.  The middle panel shows the stellar distribution superimposed upon
the total $^{13}$CO emission, while the left panel shows the position-velocity diagram
in $^{13}$CO over the same right ascension range as shown in the other panels.

As Figure \ref{fig:w13bigstars} shows, the ONC is characterized by a dense, oval 
concentration of stars with a length (in the long dimension) of about 0.2 degree 
$\sim 1.6$~pc at a distance of 470 pc.  Beyond this, the extended distribution 
becomes much more filamentary.  The elongated main body of stars 
appears to have a somewhat different position angle on the sky than the extended filament.  
These features had been recognized and characterized quantitatively
by HH98.  Fitting the stellar surface
densities (optical and infrared sources) with ellipses, HH98 found an ellipticity of
roughly 0.3 on scales (semi-major axes) less than 216 arcsec $\sim 0.5$~pc to 0.5 on a scale 
(semi-major axis) of 532 arcsec $\sim 1.2$~pc.  They also found evidence for a slight shift in
position angle of the major axis from slightly west of north on small scales to slightly
east of north on the largest scale considered.  These features appear consistent
with recent Spitzer imaging (S.T. Megeath, personal communication).  The larger scale view 
provided by the Carpenter \etal (2001) sources suggests that 
it might be better to view the ONC as a dense, moderately elongated core 
embedded in a larger-scale filamentary stellar distribution rather than an object
having a smoothly-varying ellipticity as a function of radius. 

Figure \ref{fig:w13bigstars} shows that the gas exhibits spatially-coherent
kinematic structure
in the vicinity of the ONC.  In particular, as noted above, there is clearly a larger
velocity dispersion in the gas near the ONC center.  Moreover, there are
distinct velocity systems within $\pm 0.2$~degrees of
the ONC center ($\sim \pm 1.6$~pc at a distance of 470 pc), with a distinct
hook-shaped pattern in the northern arm of the filament with a high velocity gradient
running from about $11 \kms$ at DEC $\sim -5$ to about $7 - 8 \kms$ at DEC $\sim -5.5$
(Bally \etal 1987; Wilson \etal 2005).
This behavior may suggest gravitational acceleration toward the cluster center 
(see \S 5.2).
We will discuss the implications of this structure in greater 
detail in a subsequent paper which will include
stellar radial velocities (Furesz \etal 2006).

In addition to the concentration of
stars near the dense filament, there is also a spatially-extended population
of potential members.  Many of these objects are likely to be
associated pre-main sequence stars.
Rebull \etal (2000) performed optical photometry of the fields around the main
concentration of the ONC, and demonstrated that many of these stars are pre-main sequence
members with ages $\log (t/yrs) = 6.0 \pm 0.4$, consistent with but possibly a bit older than
the stars in the central Trapezium region studied by Hillenbrand (1997).  Moreover,
Rebull \etal showed that many of these stars have ultraviolet excesses consistent with
disk accretion.  In addition, the results of both Carpenter \etal (2001) and the extended
Spitzer Space Telescope study of the region (S.T. Megeath, personal communication)
show that there are significant numbers of objects with disk-like infrared excess
emission.  Thus, the story of star formation in the region must include a consideration
of the distributed population as well as that of the ``cluster'' stars.

The significant kinematic substructure in the Orion A cloud, the
large velocity gradients and dispersions in the gas near the
ONC, and the clear extension of the ONC along the dense filament all suggest that
a dynamical model might be appropriate for understanding the observed
structure.  The overall velocity gradient in the cloud suggests that rotation
might be an important factor (Kutner \etal 1977).  
In the following section we construct an extremely simple
rotating cloud model which highlights the potential role of gravity in dynamically producing
a roughly similar structure.

\section{Model}

To demonstrate how gravitational collapse might explain the morphology of
Orion A and produce a potential protocluster gas concentration,
we construct a simple two-dimensional, self-gravitating sheet model
similar to those in BH04.
The choice of sheet geometry is motivated by a desire to construct
an initial simple model with as few free parameters as possible.
It is also consistent with our proposal that local molecular
clouds are formed by material swept up by large-scale flows
(Ballesteros-Paredes, Hartmann, \& Vazquez-Semadeni 1999; 
Hartmann, Ballesteros-Paredes, \& Bergin 2001 $=$ HBB01; Bergin \etal 2004; BH04).

BH04 showed that an initially elongated cloud will tend to collapse to a more filamentary
distribution, with a tendency to form high-density regions near the ends of
the cloud, and argued that many regions (such as Orion) showed a tendency to
have clusters form near the ends of the clouds. 
BH04 also showed that an elliptical sheet with a smooth surface density 
gradient along the major axis would produce a ``V-shaped'' cloud with a mass concentration
(potential protocluster gas) at the apex of the ``V'',  qualitatively similar to what is observed
in Orion A.  We now introduce a few refinements to this simple model to produce
an even closer resemblance to observations.

As in BH04, the numerical calculations are performed on a two-dimensional Eulerian, Cartesian
grid. The full computational region with dimension $2 \times L$ is represented
by a grid, composed of $N \times N$ grid cells, equally spaced in both directions.
Under the assumption of isothermality, the relevant differential equations to be
integrated are the  hydrodynamical continuity and momentum equations:

\begin{eqnarray}
\frac{\partial \Sigma}{\partial t} + \vec{\nabla} \cdot (\Sigma \vec{v}) & = & 0 \\
\frac{\partial \vec{v}}{\partial t} + (\vec{v} \cdot \vec{\nabla}) \vec{v} & = &
-\frac{\vec{\nabla} P}{\Sigma}- \vec{\nabla} \Phi \nonumber
\end{eqnarray}

\noindent where $\Sigma (\vec{x})$, $P(\vec{x})$ and $\vec{v}(\vec{x})$ are the
gas surface density, pressure and two-dimensional velocity vector at position
$\vec{x}$, respectively. The gravitational potential $\Phi$ is determined, solving Poisson's
equation in the equatorial plane (Binney \& Tremaine 1987)

\begin{equation}
\nabla^2 \Phi = 4 \pi G \Sigma
\end{equation}

\noindent with G the gravitational constant. The isothermal equation of state

\begin{equation}
P = c_s^2 \Sigma
\end{equation}

\noindent 
determines the pressure for a given surface density $\Sigma$ and sound speed $c_s$.

This set of equations is integrated numerically by means of an explicit finite
second-order van Leer difference scheme including operator splitting and
monotonic transport as tested and described in details in Burkert \& Bodenheimer (1993,1996).
In order to suppress numerical instabilities, an
artificial viscosity of the type described by Colella \& Woodward
(1984) is added (Burkert et al. 1997).

The Poisson equation is integrated on
the grid under the assumption that there is no matter outside of the computational region.
As we are focusing here on the gravitationally unstable sheets that collapse
towards the center of the region, outflow of gas beyond the outer boundaries can
be neglected. Therefore the outflow velocities at the outer boundary
are set to zero and a negligible pressure gradient is assumed. Most calculations were
typically performed with $256^2$ grid cells of size $\Delta = 2L/N$,
where $2L$ is the largest dimension of the rectangular computational region.
Test calculations with N=128 and N=512 did not result in significant differences.
In these calculations the code units were set such that $G = 1$.
In this paper we will focus on our favorite model with the following 
density distribution in code units:

\begin{eqnarray}
\Sigma (R) & = & \Sigma_0 \times g(x) \,\,  {\rm for} \,\,  R \leq 0.5 L\,; \nonumber \\
\Sigma (R) & = & \Sigma_0 \times g(x) \exp \left ( - \frac{(R/L-0.5)^2}{0.09} 
\right){} \\
           &   &  {}{\rm for} \,\,  0.5 L \leq R \leq 0.8 L\,; \nonumber \\
\Sigma(R)  & = & 10^{-3} \Sigma_0  \,\, {\rm for}\,\, R > L\,. \nonumber 
\end{eqnarray}

\noindent with

\begin{equation}
g(x)=0.065 + 2.17 \ \ \frac{x-L}{2 L}
\end{equation}

\noindent and

\begin{equation}
R \equiv (x^2 + \epsilon^2 y^2).
\end{equation}
Here $\epsilon =0.4$ is the initial ellipticity of the sheet.
We adopted a long axis of L=0.8 and a surface density $\Sigma_0 = 1.125$ which
gives to a total mass of $M = 1$. Finally, we introduced some initial solid-body rotation,  

\begin{eqnarray}
v_x (x,y) & = & - \Omega \, y \nonumber \\
v_y (x,y) & = & \Omega \, x \,.
\end{eqnarray}

\noindent with $\Omega = 2.5$ for this particular model.
The sound speed was set to $c_s = 0.1$ which makes the sheet
highly Jeans unstable.

\begin{figure}
\vspace*{-15mm}
\epsscale{0.8}
\plotone{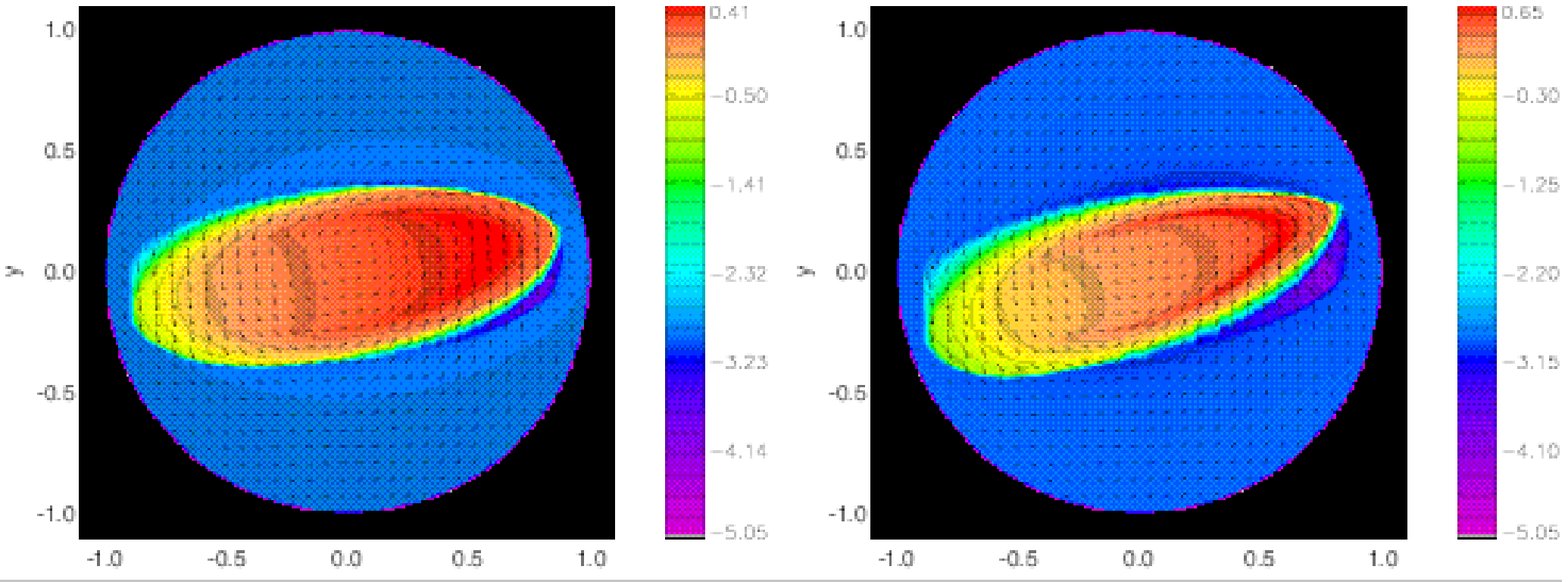}
\plotone{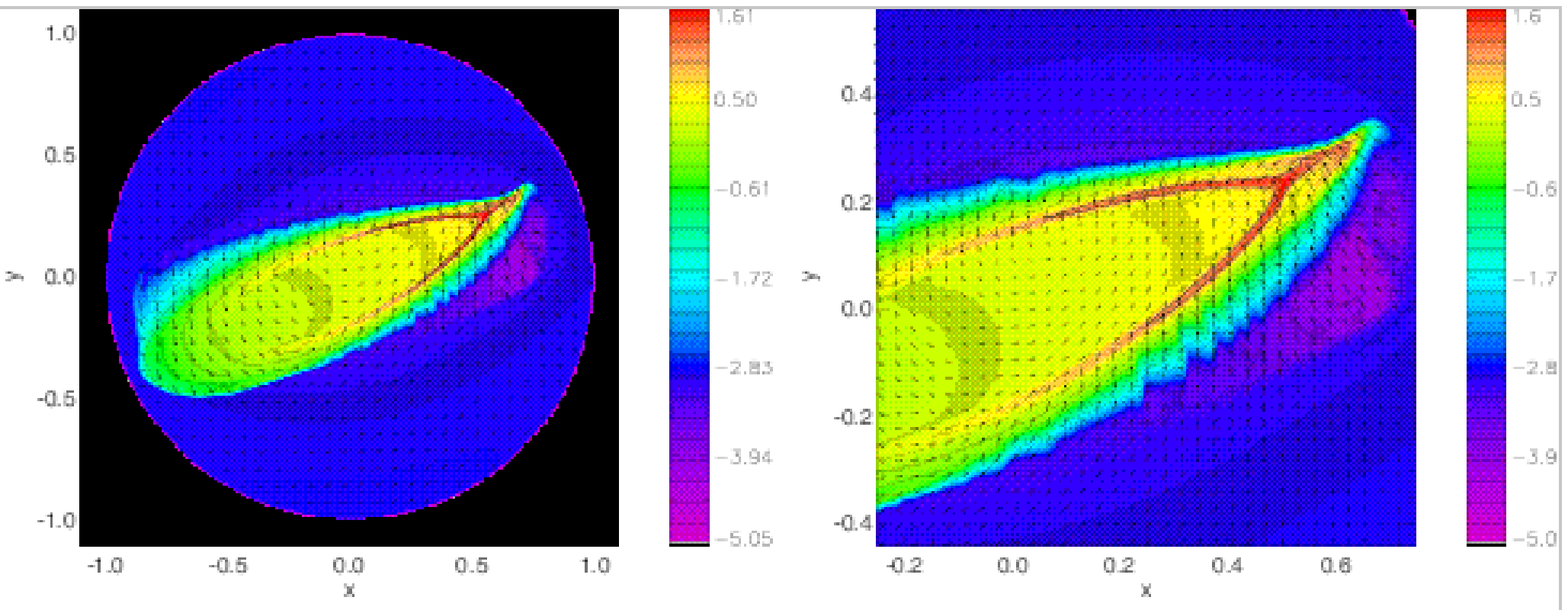}
\plotone{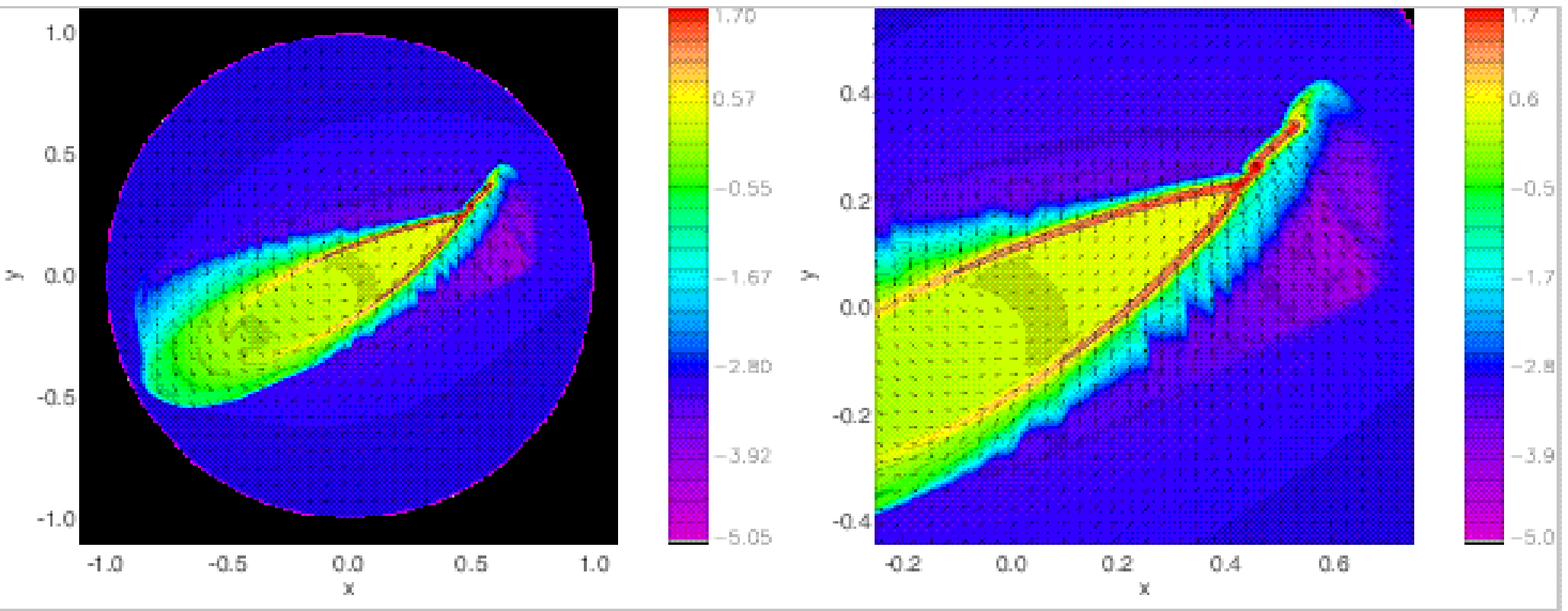}
\caption{Collapse of the elliptical rotating sheet described in the text, showing
the surface density distribution along with velocity vectors.  The color bar
indicates logarithmic values.
A time sequence is shown in the upper left and upper right panels, followed by
the middle left and lower left panels; the right middle and lower panels zoom
in on the inner regions shown in the corresponding left middle and right
panels.  The snapshots correspond to times in code units of 0.075, 0.125,
0.175, and 0.209.  The maximum velocity vector in each panel 
is 2.35, 2.82, 3.72, and 4.84 in code units, respectively, while the corresponding mass weighted
velocities are 1.07, 1.2, 1.5, and 1.7.  Note the
formation of a V-shaped structure with a filament at the upper right.
}
\label{fig:ellipse1}
\epsscale{1.0}
\end{figure}

The upper left panel of Figure \ref{fig:ellipse1} 
is a snapshot at a very early time and thus illustrates essentially the 
initial surface density distribution.  Basically, the initial model is 
an elliptical sheet with an overall smooth surface density gradient
running in the direction of the major axis and a turndown of the surface
density near the edge of the sheet.  

\begin{figure}
\plotone{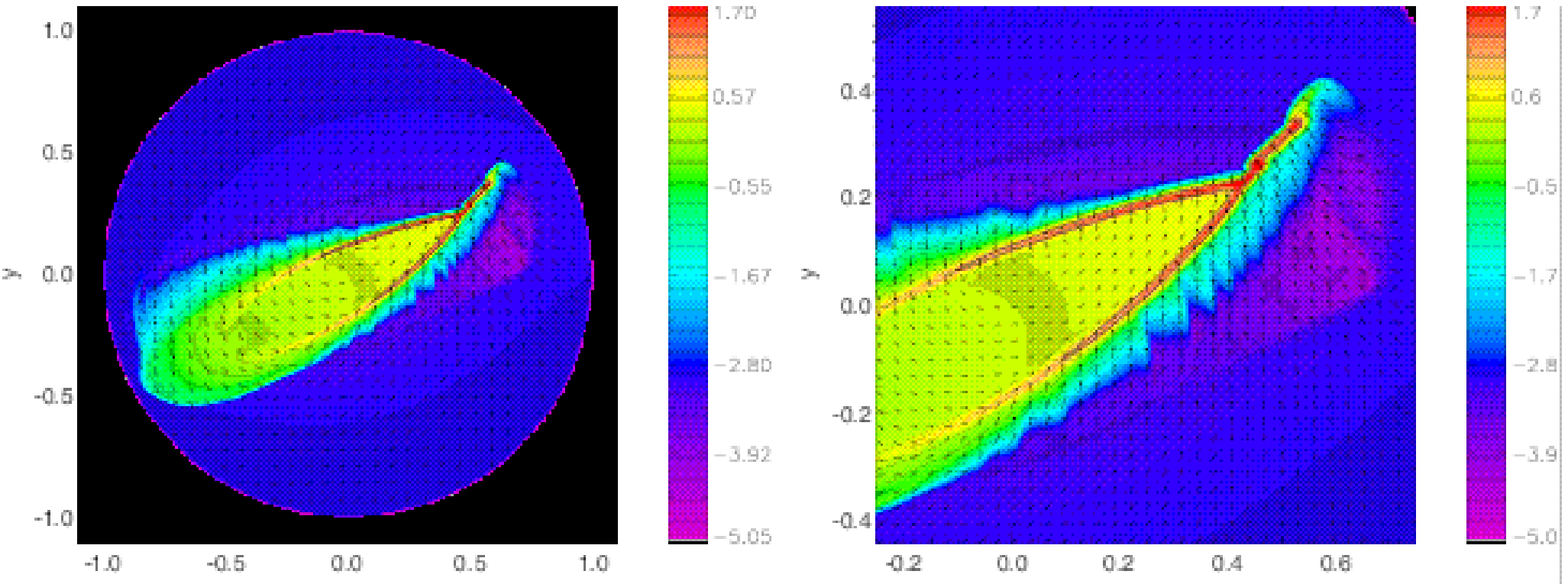}
\plotone{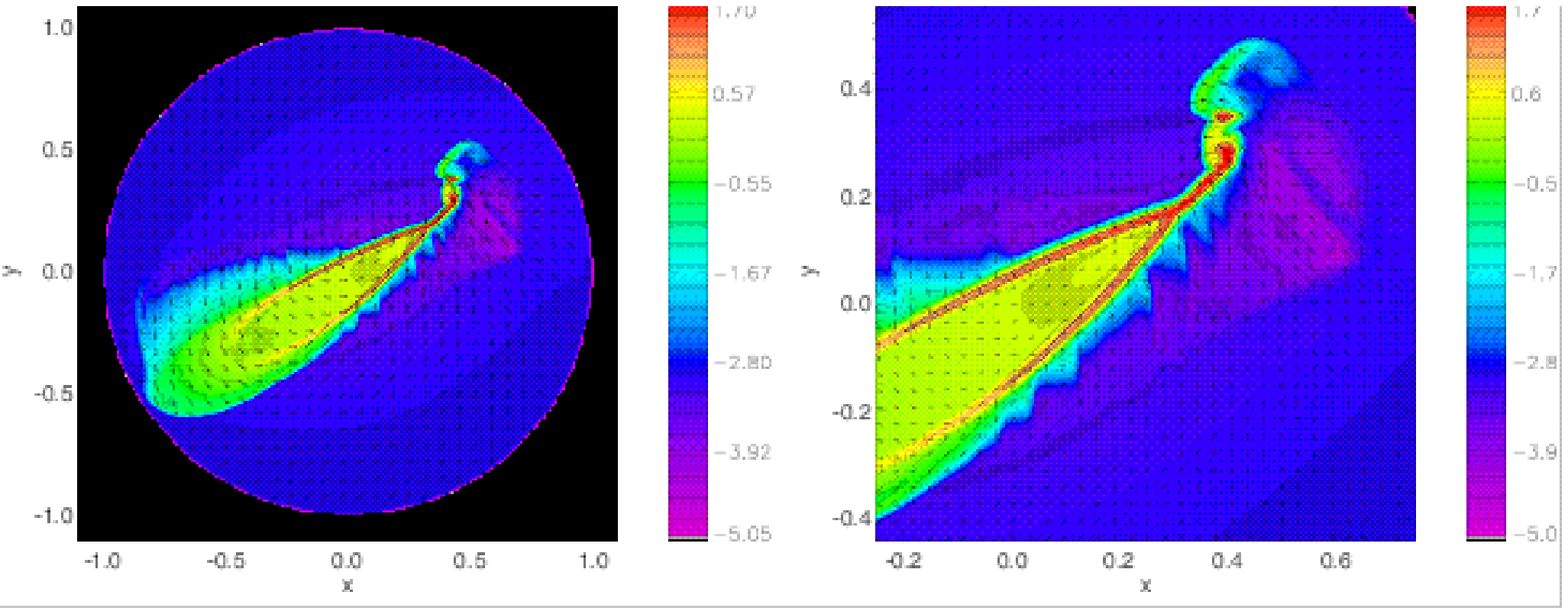}
\caption{Final stages of evolution of the model in Figure \ref{fig:ellipse1},
with the right-hand panels zooming in on the inner structure of the corresponding
left-hand panels.  The upper panels repeat the lower panels of Figure \ref{fig:ellipse1}
at a time t= 0.209 for comparison with the final snapshot at t = 0.25.
The final panels have maximum velocities of 10, with a mass-weighted velocity of 2.1
in code units. 
Note the development of something like the integral-shaped filament in the final
stage, with denser concentrations that could represent the collection of protocluster
gas (the details of these fragments are not reliable because of limited resolution).
}
\label{fig:ellipse2}
\end{figure}

Figures \ref{fig:ellipse1} and \ref{fig:ellipse2} show the time evolution of the model.
The overall density gradient produces denser structures at one end of the cloud, which
collapse faster and eventually result in 
a narrow filament.  At the wider, lower-density
end of the cloud, the collapse produces two ridges in the form of a ``V'' shape,
resulting from the non-linear gravitational acceleration that naturally occurs in
finite sheets (BH04).  The rotation of the cloud leads to a twist of the narrow
filament relative to the ``V-shaped'' lower cloud (last panels in Figure \ref{fig:ellipse2}),
and to the development of curvature in the filament reminiscent of the integral-shaped
filament.
If there were no rotation
in our model, the assumed initial conditions would result in a straight
filament aligned along the symmetry axis of the V-shaped cloud, with no
integral-shaped filament.

Dense lumps of gas are produced in the filament, which could represent
the collection of protocluster gas that could collapse and fragment into
a star cluster.  BH04 already showed the propensity of collapse
of elongated sheets to produce dense gas concentrations near the ends of filaments;
we regard this as a general result, although the details of this fragmentation (number,
mass, size) are unreliable due to limited resolution.  The turndown of density
at the edge of the original sheet results in the dense concentrations of gas lying
inside of the end of the filamentary gas, rather than right at the end of the filament
as in BH04.  The relatively large rotation of the cloud 
and the gravitational acceleration toward the narrow end
prevent the lower part of the cloud from collapsing significantly.  

The simulation does not develop small fragments that might form individual
subclusters (our resolution is far too low to examine individual star formation).  
As we found in BH04, local linear perturbations 
greater than a Jeans mass do not grow faster than the overall collapse of a sheet.  
There must be non-linear, small-scale
perturbations for stars to fragment out of this sheet, as might be produced
initially through instabilities a cloud-forming shock front
(e.g., Audit \& Hennebelle 2005; Heitsch \etal 2005, 2006; Vazquez-Semadeni \etal 2006).  

Figure \ref{fig:pv} shows position-velocity diagrams along the x and
y axes for the final configuration that corresponds to the lower panels of Figure 
\ref{fig:ellipse2}. The position-velocity distributions have been smoothed with
a top-hat filter of width $\Delta x, y = 0.025$ and $\Delta v = 0.5$.
The left panel shows the logarithm of the gas mass that moves with
y-velocity $v_y$ at position x.  The two kinematically distinct ridges of mass for
$x < 0$, corresponding to the infall from either side of the collapsing sheet.
For $0 \lesssim x \lesssim 0.25$ a third ridge is present which is the result of infall
within the central regions of the sheet toward the apex of the V.  For 
$0.3 \lesssim x \lesssim 0.4$ complex kinematic structure arises with a very large velocity
dispersion. This is a result of the gravitational acceleration of gas
in the region of influence of the dense gas lumps.
The details of this structure and the exact magnitude of these velocities
must be regarded as uncertain because of the limited resolution of the model.

The right panel of Figure \ref{fig:pv} shows somewhat similar behavior, except that
the infall in the lower part of the cloud has only a small component in the x
direction and therefore the two ridges of emission do not appear.

\section{Comparison with Orion A}

\begin{figure}
\centering
\plotone{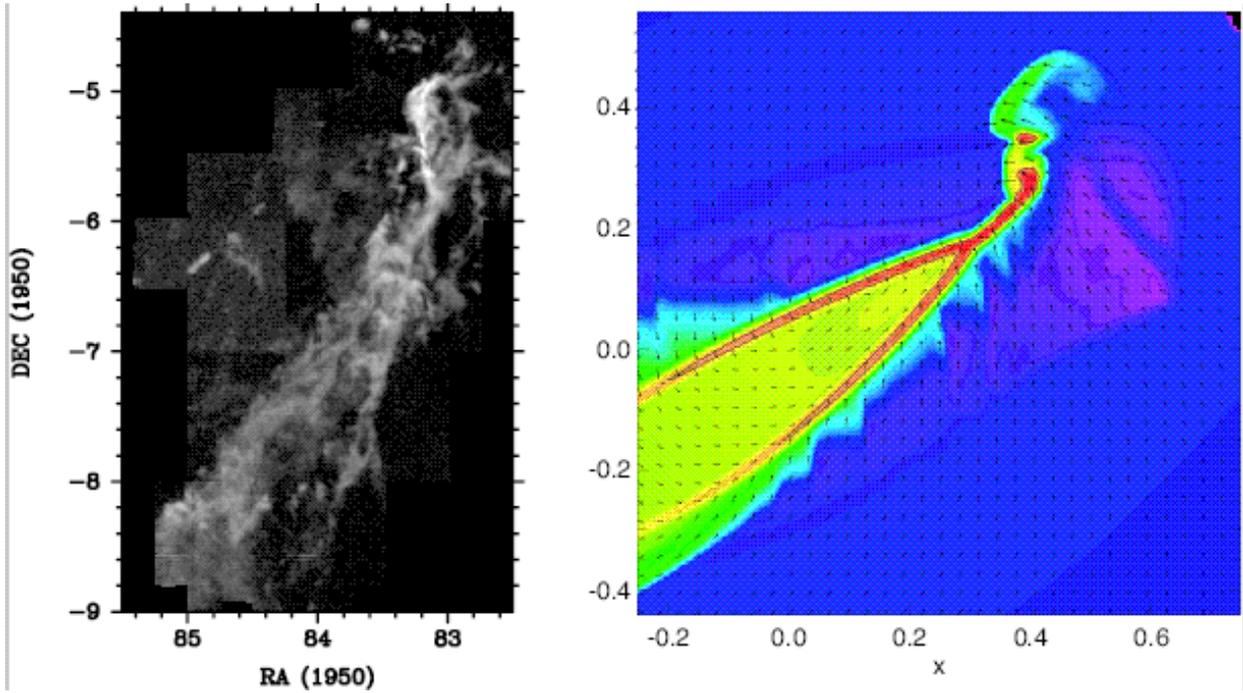}
\caption{A side-by side comparison of the $^{13}$CO emission map from the
left panel of Figure \ref{fig:w13big} with the final state of the model in
the bottom right panel of Figure \ref{fig:ellipse2}, using the scaling
discussed in the text.
}
\label{fig:oncpic1}
\end{figure}

Figure \ref{fig:oncpic1} shows a side-by-side comparison of the Orion A 
spatial distribution of $^{13}$CO emission from Figure \ref{fig:w13big}.
The model has been scaled so that one code unit of distance is $\sim 40$~pc,
a value we use in the discussion of kinematics (see below).
The qualitative resemblance of the model to the observations is clear;
the cloud narrows from an approximately ``V-shaped'' structure to an integral-shaped
filament with one dense clump and a lower density clump near the equivalent position
of the ONC in Orion A.  Interestingly, {\em Spitzer Space Telescope} observations
of the lower Orion A cloud provide evidence for an enhanced population
of young stars in the north-east ridge of the ``V'' (S.T. Megeath, personal
communication), which suggests that, as in the model, this structure really is
one of high density rather than simply high line-of-sight emission.

The model is clearly not a perfect reproduction of the Orion A cloud, which has
additional complex structure.  For example, the southwest ridge of Orion A
does not seem to be as dense as in this model.  Putting in lower densities
in the bottom half of the initial ellipse would presumably improve the agreement
with observations, but the present simple model suffices to illustrate the general
idea.  In addition, the upper part of the integral-shaped filament may not contain
sufficient mass in comparison with the real object.  This may be due in part to
resolution effects, or simply that we need a somewhat different initial density
distribution.  There is also more extended, lower-density gas seen in $^{12}$CO which
the model does not address; this would presumably require an initially more extended,
low-density envelope.  

As the motions are generated by gravity in a cloud with angular
momentum, kinematics can provide an important test of the model.
However, it is difficult to make direct test for this two-dimensional model,
as we do not know the inclination to the line of sight and the
real object is three-dimensional.  Nevertheless, we calculate scaled velocities
to provide some indication of what might be observed.

The code unit of velocity corresponds to physical units
\be
v(unit) = (G M(unit)/(R(unit)))^{1/2}\,,
\en
where $M(unit)$ is the physical mass that corresponds to 1 code mass unit
and $R(unit)$ is the physical length that corresponds to one code length unit.
We are basically modelling what Wilson \etal (2005) called ``region 2'' in
the Orion A cloud, which ranges from galactic longitudes of $208^{\circ}$
to $213^{\circ}$ at a galactic latitude $\sim -20^{\circ}$; hence we take
the total length of the cloud to be $5^{\circ} \times \cos(20) \sim 39$~pc
at a distance of 470 pc.
Wilson \etal (2005) give the total mass in region 2 as $M = 7 \times 10^4 \msun$.  
At the end of our simulation the region that corresponds to region 2
has a length of $\sim 1$ code units and contains a total mass of order 0.6 mass units.
Thus a length unit corresponds to $R(unit) \approx 39$~pc and a mass unit to
$M(unit) \approx 1.1 \times 10^5 \msun$ (which is approximately the total
Orion A cloud mass estimated by Wilson \etal 2005). 
Then the velocity code unit becomes $v(unit) \sim 3.5 \kms$.

\begin{figure}
\plottwo{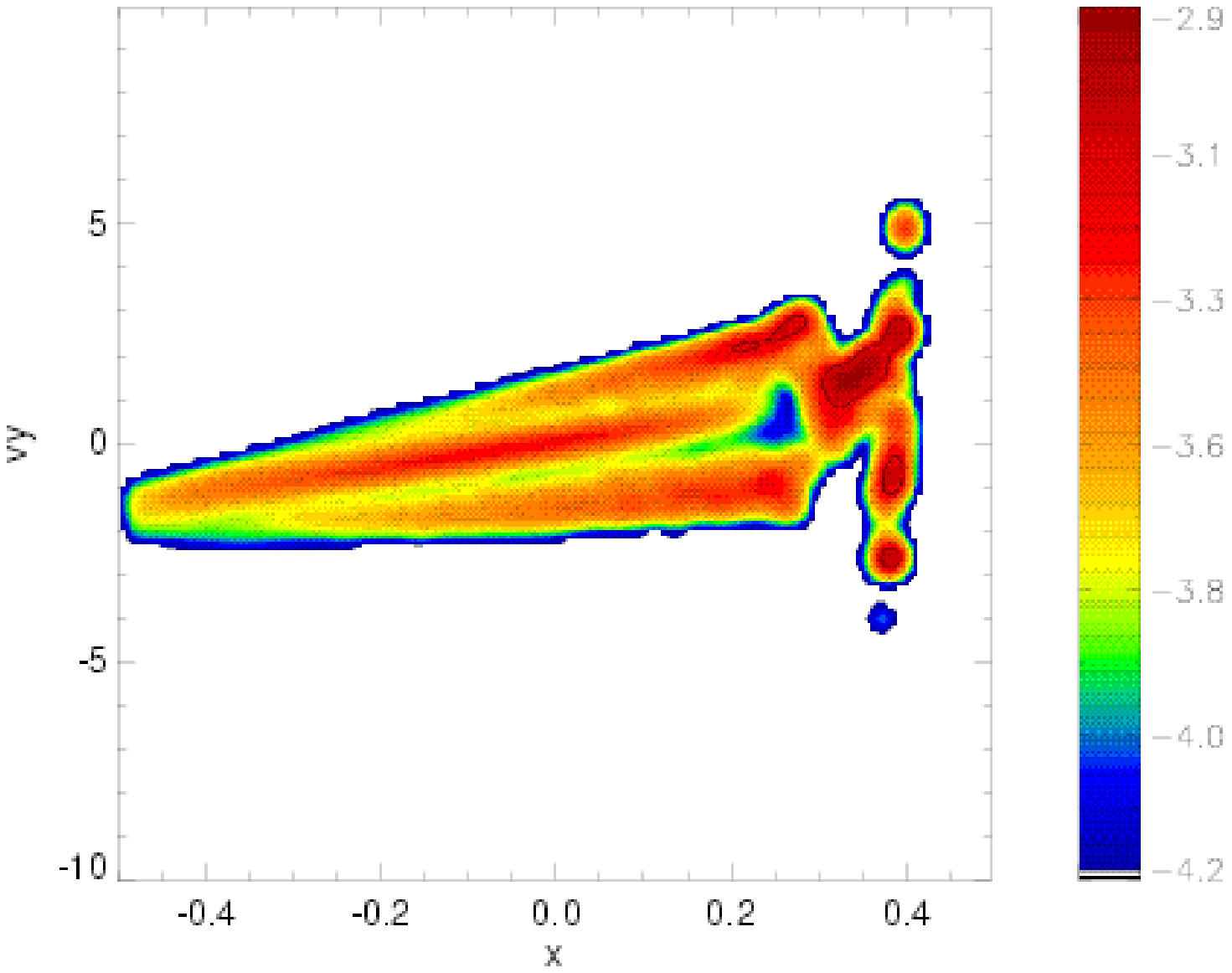}{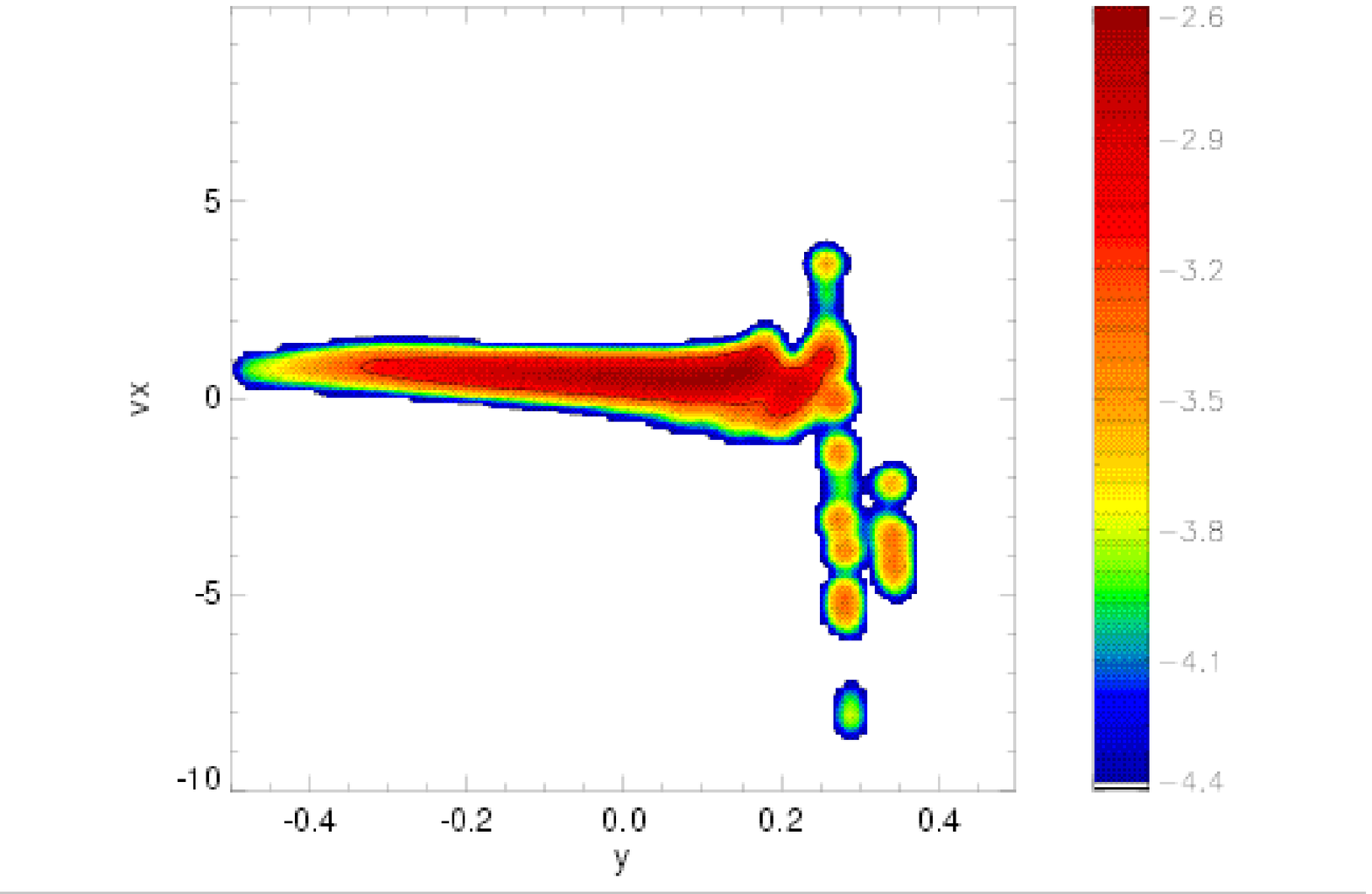}
\caption{Mass-weighted position-velocity diagrams for the last snapshot of the model
shown in Figure \ref{fig:ellipse2} (see text)}
\label{fig:pv}
\end{figure}

Applying this scaling to Figure \ref{fig:pv}, we find the following results.
In the left panel (velocities in the y-direction), the ``filaments'' 
at $x < 0.2$ are spaced by approximately 1.5 to 2 code units, or $\sim 5 - 7 \kms$.
The ``middle filament'' in position-velocity space has an overall velocity shift from
$-0.5 < x < 0.2$ of slightly less than 2 units, or about $6 \kms$.
The observed values of occasional velocity splitting and overall velocity
gradient (right panel of Figure \ref{fig:w13big} are about 1/2 to 2/3 as large
as these values, so the agreement is reasonable, especially considering 
possible projection effects; even ignoring the inclination of the sheet to
the line of sight, the velocities in the x-direction (right panel) are much
smaller than observed, because of the strongly anisotropic nature of the
velocity field.  We also note that some of the velocity splitting in Orion A
may be due to stellar-wind driven bubbles (see \S 5.5) which is not part
of this simple model.

The model velocity dispersions near the apex of the V and the 
filament are much larger than the radial velocity dispersions observed in
the similar regions in Orion A.  At $x \sim 0.25$, the velocity width in the left panel is
about 4 units, or $\sim 14 \kms$, while the radial velocity width of the
$^{13}$CO emission is closer to $4 \kms$.  Again, the x velocity shows very little
velocity dispersion.
This might indicate an additional tilt of the gas structure
of Orion A in the third (z) direction that cannot be modeled by our 2-dimensional
simulations.  Both projections show very large velocity dispersions near the
dense concentration(s) in the integral-shaped filament, of order 6 code units or
$\sim 20 \kms$ that are much larger than observed.  However, it should be emphasized
that near the tip of the V and the filament the results are uncertain because of
limited resolution, which could have a major effect on the calculated velocities.
As indicators of this, the separate ``blobs'' in position-velocity space seen
in Figure \ref{fig:pv} represent just a few grid cells each at most.  In addition,
we note that the maximum velocities in the computation essentially double from
the second-last snapshot at $t=0.209$ to $t = 0.25$, suggesting great sensitivity
to the chosen time of comparison - probably also a reflection of limited resolution.
Finally, the gravitational potential of the stars in the ONC should also be taken
into account in an improved model.

The rotational motion making the integral-shaped filament is especially apparent
in the x-velocity (right panel of Figure \ref{fig:pv}), where the generally
positive velocity components quickly reverse to negative velocities at $x > 0.25$.
This raises the question as to whether the hook-shaped velocity structure in the
northern half of the real filament (Figure \ref{fig:w13bigstars}) might be a signature
of the rotational motion, accelerated by gravitational collapse.  To be consistent
with the model, one would have to assume that the plane of the initial sheet was
tilted so that its western side is closer to Earth.

We conclude that our model velocities are at least consistent with observed
gas radial velocities in the sense that they can be as large or larger
than observed.  It is difficult to say more than this, given the unknown projection effects
and the uncertainty in the velocity field near the poorly-resolved densest regions.
Higher-resolution, 3-dimensional simulations
are required to explore this issue further.

\section{Discussion}

We have shown that a gravitationally-collapsing cloud with smooth,
simple properties can evolve into a more complex structure qualitatively
similar to that of Orion A (Figure \ref{fig:oncpic1}).  
Although our specific model parameters are not unique, the general requirements
are plausible:

(1) The initial molecular cloud was elongated.  Flow-produced clouds will naturally
tend to be flattened, and the general case would be for one lateral dimension
to be longer than the other. 

(2) The initial internal velocities were sufficiently small to allow 
collapse, or they dissipated sufficiently quickly to allow collapse.

(3) The cloud was formed with angular momentum.  

(4) The elongated cloud initially had more mass at one end than the other and
had a turndown of density near its edge.

Our model provides a natural explanation for the formation
of the narrow integral-shaped filament without requiring specific focusing flows driven by an
unknown mechanism. The condensations which we find the S-shaped filament could
form a cluster of stars like the ONC. The large velocity gradient observed in this region
in our simulations results from gravitationally accelerated gas when falling into the deep
potential well generated by the condensation, coupled with rotation.
The collapse model 
is dynamically plausible; all that is required is to assume that the stellar energy
input, particularly that of the central O star has not yet disrupted the dense gas. 
In contrast, equilibrium or quasi-equilibrium models 
must explain how to support a very massive filament against its self-gravity
and how to maintain the observed massive, supersonically-moving clumps of gas
without either overshooting (blowing the region apart) or undershooting 
(gravitational collapse, as in our picture).

The model predicts that there should be some measurable rotation in the plane of the sky;
although the model velocities are uncertain in this region, they could easily be
several $\kms$.  The proper motions of stars in the ONC should reflect
this motion to some extent, particularly the stars at either end where the cluster
is most filamentary.  Unfortunately, the methods used
by Jones \& Walker (1988) to measure proper motions were such as to eliminate any
effect of rotation (or expansion; e.g., Scally, Clarke, \& McCaughrean 2005).
Future proper motion measurements should provide a direct test of the model. 

\subsection{Ages and structure}

Another perhaps initially surprising feature of this model is the rapidity of the collapse.
The physical time of the code time units is $(R^3/GM)^{1/2}$, which, for the
parameters adopted in \S 4, is $t(unit) = 6.7$~Myr.  Thus, the time taken for the
collapse of the original ellipse, with an initial semi-major axis of 24 pc and a 
semi-minor axis of 4.8 pc, is only 1.7 Myr.  Obviously this is not a very precise
estimate, as it depends crucially on just what the original form of the cloud
was, and perhaps on the timescale of dissipation of local turbulence, but it is
clearly consistent with the typical estimated ages of 1-3 Myr for stars in the 
region (Hillenbrand 1997; Rebull \etal 2000).

As shown in the right-hand panel of Figure \ref{fig:w13big},
there are clearly lumps of gas correlated in space and in velocity; these regions
cannot have experienced more than about one crossing or collapse
time; otherwise the velocity structure would be erased as the infalling gas shocks
and dissipates its kinetic energy. The stars would tend to pass through the shocked gas,
reducing the spatial correlation of stars with dense gas.  All of these features
suggest that Orion A is dynamically young, in agreement with our model. 

\subsection{Dynamics and Formation of the ONC}

Our model suggests that the gas for the ONC was assembled by
large-scale gravitational collapse.  In contrast,
Tan, Krumholz, \& McKee (2006; TKM06) have argued for a picture in which 
rich star clusters take several dynamical times to form, are quasi-equilibrium 
structures during formation, and thus initial conditions are not very important.
In particular, TKM06 discuss the ONC and argue that it is several crossing times
old, and argue that its smooth structure supports their quasi-equilibrium picture 
(see also Scally \& Clarke 2002).  
However, it is difficult to see how to maintain an equilibrium
condition in a system with many thermal Jeans masses, rapidly dissipating turbulence,
and the destructive effects of stellar energy input; as Bonnell \& Bate (2006) argue,
it is much easier to undershoot or overshoot than maintain a balance.  

In a subsequent paper (Furesz \etal 2006) we will discuss evidence
from the stellar kinematics
suggesting that the ONC is not in a dynamical equilibrium. 
Here we simply make two points.  First, the spatially-coherent kinematic
substructure seen in the vicinity of the ONC (Figures
\ref{fig:w13big} and \ref{fig:w13bigstars}) does not obviously suggest equilibrium
conditions; the large velocity gradient in the northern end of the integral-shaped
filament might be produced by infalling gravitationally-accelerated gas
(which would not be surprising, as the mass of the stars in the cluster
provides additional acceleration toward the center.
Second, the stellar distribution is clearly filamentary on radial scales
$r \gtrsim 1$~pc, not typical of a relaxed system.

\subsection{Bound Clusters?}

Early discussions of cluster formation (e.g., Lada, Margulis, \& Dearborn 1984; Mathieu 1983)
assumed initial conditions such that the stars had virial kinetic energies and
then considered how efficient star formation must be to end up with a bound (open)
cluster. More recently, Adams (2000) studied the dependence
of bound cluster formation on its anisotropy.
Geyer \& Burkert (2001) showed that a cluster which is already in 
virial equilibrium when the remaining gas is ejected will only remain gravitationally bound
only if more than $50 \%$ of the gas was converted into stars.
Gas expulsion from a collapsing star cluster with initially small velocity dispersion
could however produce a bound cluster even for low star formation 
efficiencies of order $10 \%$ (see also Lada \etal 1984).
In the simulations discussed by Bonnell, Bate, \& Vine (2003), 
which assumed an initial turbulent velocity field in the gas with 
dispersion comparable to the virial value, 
the turbulence quickly dissipated as stars formed and the cluster gas began
to contract gravitationally.  Our model suggests that the infall motions
might be even more important initially; the overall collapse of the cloud,
with the consequent effect of producing a time-variable gravitational potential,
could increase the importance of violent relaxation in altering the stellar
distribution (e.g., Binney \& Tremaine 1987).
Gravitational collapse produces velocities only $2^{1/2}$ times
larger than virial values, and it is still quite possible that a bound remnant
cluster can result with even modest efficiencies of star formation depending upon
the details of the stellar spatial and velocity distributions (e.g.,
Boily \& Kroupa 2003).  Addressing these issues further will 
require more sophisticated simulations including stars as well as gas.

\subsection{Cloud dynamics}

The collapse picture naturally produces large-scale velocity gradients of
``virial'' magnitude, as seen in the north-south gradient in Orion A; 
approximate agreement between kinematic and gravitational terms as inferred from
the virial theorem is obviously not a good indicator of true virial equilibrium
(Ballesteros-Paredes 2006).  Given an initial cloud angular momentum,
collapse might also naturally result in large scale rotation near centrifugal
balance, as suggested by the radial velocity gradient observed in Orion A. 
This model also suggests that a significant fraction of the ``turbulence'' could be
gravitationally-generated.  Stellar energy
input such as outflows, photoionization, and winds will certainly produce additional
turbulence, especially in the low-density regions; but gravity should be
much more efficient in accelerating large, dense lumps of gas and stars.
Indeed, as BH04 pointed out, it is extremely difficult if not practically
impossible to avoid generating significant gravitational ``turbulence'' in
clouds of many thermal Jeans masses if there is any substructure to the cloud
(though these motions might be called more properly gravitational flows than 
turbulence).

The collapse model implies that Orion A is globally magnetically
supercritical; otherwise the magnetic fields would prevent collapse.
A similar conclusion applying to molecular clouds in general
was reached by Bertoldi \& McKee (1992) on different grounds.  
If molecular clouds are globally supercritical,
then any subcritical regions must be balanced by other, even more than average
supercritical regions.  The further implication is that
there is in general little if any magnetic
barrier to star formation, and so star formation can be ``fast'',
as required observationally (HBB01).

Global collapse would also make it easier to assemble
molecular clouds from large-scale flows in the diffuse atomic interstellar
medium in the solar neighborhood.  Bergin \etal (2004) computed one-dimensional
shock models with chemistry and showed that 
the requirement of a minimum surface density to shield the CO from the interstellar
ultraviolet radiation field might imply sweep-up timescales of
several to 20 Myr, depending upon the initial pre-shock density.  
However, if the cloud can gravitationally contract in the {\em lateral}
direction, there can be a runaway process of increasing column density on much
shorter timescales - a few Myr or less, in the case of Orion A.

\subsection{Stellar energy input}

If global gravitational collapse is a general feature of nearby
molecular clouds, why is the local star formation rate so low?  
Turbulent support is ineffective in preventing collapse in dense
regions (e.g., Stone, Ostriker, \& Gammie 1998; Mac Low et al. 1998;
Heitsch, Mac Low, \& Klessen 2001)
unless driven, and even then it is difficult to inject the energy into
the densest fragments most in need of support 
(Balsara 1996; Elmegreen 1999; Heitsch \etal 2005, 2006).
The high-mass stars in Orion A will almost certainly disrupt at least the
northern end of the ONC; $\theta^1$~C Ori is already photodissociating and
photoionizing the filament gas at a rapid rate (O'Dell 2001),
it will probably become a supernova and disrupt much of the remaining cloud
in 3-5 Myr.  The efficiency of low-mass stars in blowing away material from
the southern end of the cloud might be significant, though this is less
certain (e.g., Matzner \& McKee 2000).

The possibility that much or all of the cloud motions and structure discussed
here are due to stellar energy input is difficult to reject conclusively.
Bally \etal (1987) suggested that the entire cloud might be shaped by
pressure from the Orion superbubble, and Wilson \etal (2005) suggested that
the hook-shaped velocity structure might also be due to pressure from
the stellar winds of the Orion 1b subgroup.  Heyer \etal (1992) noted the
presence of several bubble-like structures in $^{12}$CO (see also Bally 1989)
which might be driven by stellar winds.  However, Heyer \etal also
noted that the energetics of the observed bubbles were probably
not sufficient to disrupt the L1641 (lower) cloud region.
The bubbles are less apparent in the $^{13}$CO map, which traces 
denser gas than $^{12}$CO (cf. Wilson \etal 2005).  Moreover, it is hard
to see how the structure and the coherent streaming
velocities of the extremely dense and massive integral-shaped
filament can be created in this way.  Indeed, it is not obvious
from the $^{13}$CO intensity map alone just where the ONC and the central
ionizing O star reside, as one might expect if stellar energy input were
dominant.  Similarly, the northwest ridge or the lower region
of the Orion A cloud exhibits the highest concentration of young stars 
(S.T. Megeath, personal communication) which is not an obvious result of 
driving by the lower-mass stars in this region.  For these reasons 
we argue that driving by stellar energy input, while significant in 
making substructure within the cloud, is not the major influence
shaping the global structure of Orion A. 

\section{Conclusions}

We have developed a simple model of a gravitationally collapsing molecular cloud
which accounts the overall morphology of the Orion A cloud and suggests
the formation of a mass concentration in a location similar to that
of the ONC.  Further kinematic studies, especially stellar proper
motions, can be used to test this picture.  Our model implies that star cluster
formation at least in some cases is a dynamic, complex process, with gravitational collapse playing
an essential role in assembling the protocluster gas.  Future simulations in three
dimensions including stars as well as gas will be needed to test new and improved
kinematic measurements of the stellar population, providing additional insight
into the formation of the ONC. 

\acknowledgments

We are grateful to Bob Wilson for providing the $^{13}$CO data of Orion A, to 
John Carpenter for providing the stellar positions of the infrared sources
found in his survey, and to Fabian Heitsch with IDL help and comments on the
manuscript.  

\clearpage

\begin{thebibliography}

\bibitem[Adams(2000)]{2000ApJ...542....964A} Adams, F.C.\ 2000, \apj, 542, 964

\bibitem[Audit \& Hennebelle(2005)]{2005A&A...433....1A} Audit, E., \& 
Hennebelle, P.\ 2005, \aap, 433, 1

\bibitem[Ballesteros-Paredes (2006)]{javier06} Ballesteros-Paredes, J. 2006,
astro-ph/0606071 

\bibitem[Ballesteros-Paredes et al. (1999a)]{javier99} 
Ballesteros-Paredes, J., V{\'a}zquez-Semadeni, E., \& Scalo, J.\ 1999, \apj, 515, 286 
{1999ApJ...527..285B} 

\bibitem[Ballesteros-Paredes et al. (1999b)]{javier99b}
Ballesteros-Paredes, J., Hartmann, L., \& V{\'a}zquez-Semadeni, E.\ 1999, 
\apj, 527, 285

\bibitem[Ballesteros-Paredes et al. (2006)]{ppv06}
Ballesteros-Paredes, J., Klessen, R.S., Mac Low, M.-M., \& Vazquez-Semadeni, E. 2006,
in Protostars and Planets V, eds. B. Reipurth, D. Jewitt, \& K. Keil (Tucson: 
University of Arizona Press), in press (astro-ph/0603357)

\bibitem[Bally (1989)]{1989proc} Bally, J., in Proc. ESO Workshop on Low-Mass Star
Formation and Pre-Main Sequence Objects, ed. B. Reipurth (Garching: ESO), 1

\bibitem[Bally et al.(1987)]{1987ApJ...312L..45B} Bally, J., Stark, A.~A., 
Wilson, R.~W., \& Langer, W.~D.\ 1987, \apjl, 312, L45

\bibitem[Balsara(1996)]{1996ApJ...465..775B} Balsara, D.~S.\ 1996, \apj, 
465, 775

\bibitem[Bate et al.(2002)]{bbb02} Bate, M.~R., Bonnell, I.~A., \& Bromm, V.\ 2002, \mnras, 332, L65

\bibitem[Bate et al.(2003)]{bbb03} Bate, M.~R., Bonnell, I.~A., \& Bromm, V.\ 2003, \mnras, 339, 577

\bibitem[Bergin et al.(2004)]{2004ApJ...612..921B} Bergin, E.~A., Hartmann, 
L.~W., Raymond, J.~C., \& Ballesteros-Paredes, J.\ 2004, \apj, 612, 921

\bibitem[Bertoldi \& McKee(1992)]{1992ApJ...395..140B} Bertoldi, F., \& 
McKee, C.~F.\ 1992, \apj, 395, 140

\bibitem[Binney \& Tremaine (1987)]{bt87} 
Binney, J. \& Tremaine, S.\ 1987, Galactic Dynamics (Princeton Univ. Press), pp 271-273

\bibitem[Boily \& Kroupa(2003)]{2003MNRAS.338..673B} Boily, C.~M., \& 
Kroupa, P.\ 2003, \mnras, 338, 673

\bibitem[Bonnell et al. (2003)]
{bbv03} Bonnell, I.~A., Bate, M.~R., \& Vine, S.~G.\ 2003, \mnras, 343, 413

\bibitem[Bonnell et al. (2006)]
{bvb06} Bonnell, I.~A., \& Bate, M.~R.\ 2006, \mnras, in press (astro-ph/0604615)

\bibitem[Burkert (2006)]
{b06} Burkert, A. 2006, in Statistical Mechanics of Non-Extensive Systems,
ed. F. Combes \& R. Robert (Elsevier), astro-ph/0605088

\bibitem[Burkert et al. (1997)]
{bbb97} Burkert, A., Bate, M.R. \& Bodenheimer, P. 1997, MNRAS, 289, 497

\bibitem[Burkert \& Bodenheimer (1993)]
{bb93} Burkert, A. \& Bodenheimer, P.\ 1993, \mnras, 264, 798

\bibitem[Burkert \& Bodenheimer (1996)]
{bb96} Burkert, A. \& Bodenheimer, P.\ 1996, \mnras, 280, 1190

\bibitem[Burkert \& Hartmann(2004)]{2004ApJ...616..288B} Burkert, A., \& 
Hartmann, L.\ 2004, \apj, 616, 288 (BH04) 

\bibitem[Carpenter et al.(2001)]{2001AJ....121.3160C} Carpenter, J.~M., 
Hillenbrand, L.~A., \& Skrutskie, M.~F.\ 2001, \aj, 121, 3160

\bibitem[Collella \& Woodward (1984)]
{colwood} Collela, P. \& Woodward, P. 1984, J. Comput. Phys. 54, 174

\bibitem[Elmegreen(2000)]{2000ApJ...530..277E} Elmegreen, B.~G.\ 2000, 
\apj, 530, 277

\bibitem[Genzel \& Stutzki(1989)]{1989ARA&A..27...41G} Genzel, R., \& 
Stutzki, J.\ 1989, \araa, 27, 41

\bibitem[Geyer \& Burkert(2001)]{2001MNRAS..323...988G} Geyer, M.P., \& 
Burkert, A.\ 2001, \mnras, 323, 988

\bibitem[Furesz et al. (2006)]{furesz et al}
Furesz, G. \etal 2006, in preparation

\bibitem[Hartmann et al.(2001)]{2001ApJ...562..852H} Hartmann, L., 
Ballesteros-Paredes, J., \& Bergin, E.~A.\ 2001, \apj, 562, 852  (HBB01)

\bibitem[Hillenbrand(1997)]{1997AJ....113.1733H} Hillenbrand, L.~A.\ 1997, 
\aj, 113, 1733

\bibitem[Hillenbrand \& Carpenter(2000)]{2000ApJ...540..236H} Hillenbrand, 
L.~A., \& Carpenter, J.~M.\ 2000, \apj, 540, 236

\bibitem[Hillenbrand \& Hartmann(1998)]{1998ApJ...492..540H} Hillenbrand, 
L.~A., \& Hartmann, L.~W.\ 1998, \apj, 492, 540 (HH98) 

\bibitem[Heitsch et al.(2005)]{2005ApJ...633L.113H} Heitsch, F., Burkert, 
A., Hartmann, L.~W., Slyz, A.~D., \& Devriendt, J.~E.~G.\ 2005, \apjl, 633, 

\bibitem[Heitsch et al.(2006)]{2006ApJ} Heitsch, F., Burkert, 
A., Hartmann, L.~W., Slyz, A.~D., \& Devriendt, J.~E.~G.\ 2006, \apj, in press

\bibitem[Heitsch et al.(2001)]{2001ApJ...547..280H} Heitsch, F., Mac Low, 
M.-M., \& Klessen, R.~S.\ 2001, \apj, 547, 280

\bibitem[Heyer et al.(1992)]{1992ApJ...395L..99H} Heyer, M.~H., Morgan, J., 
Schloerb, F.~P., Snell, R.~L., \& Goldsmith, P.~F.\ 1992, \apjl, 395, L99

\bibitem[Jones \& Walker(1988)]{1988AJ.....95.1755J} Jones, B.~F., \& 
Walker, M.~F.\ 1988, \aj, 95, 1755

\bibitem[Klessen \& Burkert(2000)]{2000ApJS..128..287K} Klessen, R.~S., \& 
Burkert, A.\ 2000, \apjs, 128, 287

\bibitem[Klessen \& Burkert(2001)]{2001ApJ...549..386K} Klessen, R.~S., \& 
Burkert, A.\ 2001, \apj, 549, 386

\bibitem[Krumholz et al.(2005)]{2005Natur.438..332K} Krumholz, M.~R., 
McKee, C.~F., \& Klein, R.~I.\ 2005, \nat, 438, 332

\bibitem[Kutner et al.(1977)]{1977ApJ...215..521K} Kutner, M.~L., Tucker, 
K.~D., Chin, G., \& Thaddeus, P.\ 1977, \apj, 215, 521

\bibitem[Lada \& Lada(2003)]{2003ARA&A..41...57L} Lada, C.~J., \& Lada, 
E.~A.\ 2003, \araa, 41, 57

\bibitem[Lada et al.(1984)]{1984ApJ...285..141L} Lada, C.~J., Margulis, M., 
\& Dearborn, D.\ 1984, \apj, 285, 141

\bibitem[Mac Low et al.(1998)]{1998PhRvL..80.2754M} Mac Low, M.-M., 
Klessen, R.~S., Burkert, A., \& Smith, M.~D.\ 1998, Physical Review 
Letters, 80, 2754

\bibitem[Maddalena et al.(1986)]{1986ApJ...303..375M} Maddalena, R.~J., 
Morris, M., Moscowitz, J., \& Thaddeus, P.\ 1986, \apj, 303, 375

\bibitem[Mathieu(1983)]{1983ApJ...267L..97M} Mathieu, R.~D.\ 1983, \apjl, 
267, L97 

\bibitem[Matzner \& McKee(2000)]{2000ApJ...545..364M} Matzner, C.~D., \& 
McKee, C.~F.\ 2000, \apj, 545, 364

\bibitem[O'Dell(2001)]{2001ARA&A..39...99O} O'Dell, C.~R.\ 2001, \araa, 39, 99

\bibitem[Ostriker et al. (1999)]
{ogs99} Ostriker, E.~C., Gammie, C.~F., \& Stone, J.~M.\ 1999, \apj, 513, 259

\bibitem[Rebull et al.(2000)]{2000AJ....119.3026R} Rebull, L.~M., 
Hillenbrand, L.~A., Strom, S.~E., Duncan, D.~K., Patten, B.~M., Pavlovsky, 
C.~M., Makidon, R., \& Adams, M.~T.\ 2000, \aj, 119, 3026

\bibitem[Scally \& Clarke(2002)]{2002MNRAS.334..156S} Scally, A., \& 
Clarke, C.\ 2002, \mnras, 334, 156

\bibitem[Scally et al.(2005)]{2005MNRAS.358..742S} Scally, A., Clarke, C., 
\& McCaughrean, M.~J.\ 2005, \mnras, 358, 742

\bibitem[Stone et al.(1998)]{1998ApJ...508L..99S} Stone, J.~M., Ostriker, 
E.~C., \& Gammie, C.~F.\ 1998, \apjl, 508, L99

\bibitem[Tan et al.(2006)]{2006ApJ...641L.121T} Tan, J.~C., Krumholz, 
M.~R., \& McKee, C.~F.\ 2006, \apjl, 641, L121 (TKM06)

\bibitem[V{\'a}zquez-Semadeni et al.(2006)]{2006ApJ...643..245V} 
V{\'a}zquez-Semadeni, E., Ryu, D., Passot, T., Gonz{\'a}lez, R.~F., \& 
Gazol, A.\ 2006, \apj, 643, 245

\bibitem[Wilson et al.(2005)]{2005A&A...430..523W} Wilson, B.~A., Dame, 
T.~M., Masheder, M.~R.~W., \& Thaddeus, P.\ 2005, \aap, 430, 523

\end{thebibliography}
\end{document}